# Unveiling Placental Development in Circadian Rhythm-Disrupted Mice: A Photo-acoustic Imaging Study on Unstained Tissue

M. N. Cizmeciyan,[1,4] N. I. Bektas,[2] N. Derin[3], T. Denizaltı[3], A. Khoshzaban[1], M. B. Unlu[1,4], and C. Celik-Ozenci[5,6]

[1]*Department of Physics, Bogazici University, Bebek, Istanbul, 34342, Turkey*
[2]*Department of Histology and Embryology, School of Medicine, Akdeniz University, Antalya, Turkey*
[3]*Department of Biophysics, School of Medicine, Akdeniz University, Antalya, Turkey*
[4]*Faculty of Engineering, Ozyegin University, Cekmekoy, Istanbul, 34794, Turkey*
[5]*Department of Histology and Embryology, School of Medicine, Koc University, Rumelifeneri, Sariyer, Istanbul 34450, Turkey*
[6]*Koc University Research Center for Translational Medicine (KUTTAM), Koc University, Istanbul, Turkey*

**Abstract:**

*Introduction:* Circadian rhythm disruption has garnered significant attention for its adverse effects on human health, particularly in reproductive medicine and fetal well-being. Assessing pregnancy health often relies on diagnostic markers such as the labyrinth zone (LZ) proportion within the placenta. This study aimed to investigate the impact of disrupted circadian rhythms on placental health and fetal development using animal models.

*Methods and Results:* Employing unstained photo-acoustic microscopy (PAM) and hematoxylin and eosin (HE)-stained images, we found them mutually reinforcing. Our images revealed the role of MCRD on the LZ and fetus weight: a decrease in LZ area from 5.01-HE(4.25-PAM) mm$^2$ to 3.58-HE (2.62-PAM) mm$^2$ on day 16 and 6.48-HE(5.16-PAM) mm$^2$ to 4.61-HE (3.03-PAM) mm$^2$ on day 18, resulting in 0.71 times lower fetus weights. We have discriminated a decrease in the mean LZ to placenta area ratio from 64% to 47% on day 18 in mice with disrupted circadian rhythms with PAM.

*Discussion:* The study highlights the negative influence of circadian rhythm disruption on placental development and fetal well-being. Reduced LZ area and fetal weights in the MCRD group suggest compromised placental function under disrupted circadian rhythms. PAM imaging proved to be an efficient technique for assessing placental development, offering advantages over traditional staining methods. These findings contribute to understanding the mechanisms underlying circadian disruption's effects on reproductive health and fetal development, emphasizing the importance of maintaining normal circadian rhythms for optimal pregnancy outcomes. Further research is needed to explore interventions to mitigate these effects and improve pregnancy outcomes.

## 1. Introduction

Photoacoustic imaging (PAI) produces highly sensitive vasculature images in tissue due to the high absorption cross-section of red blood cells [1]. The linear relation between laser absorption by RBCs and the magnitude of the generated photoacoustic waves allows us to create stain-free, concentration-dependent images of blood vessels [2]. This unique capability distinguishes PAI from conventional imaging modalities, facilitating the precise identification of hypoxia – the underlying cause of many diseases [3]. However, while PAI holds promise for clinical applications, current implementations predominantly focus on cancer and vascular disorders, overlooking a vital topic: assessing placental health. The placental interface is crucial for meeting the metabolic demands of the developing fetus through nutrition exchange mediated by blood, and disruptions in this communication have been associated with negative fetal outcomes [4-7]. A few studies in placental PAI provided insights into the limitations and strengths but remained limited in understanding placental physiology fully. Early studies, beginning in 2015 with ex vivo placental tissue [8], progressed to accurate measurements of umbilical cord oxygenation in 2016 [9, 10]. By 2017, the first pathological case demonstrated the placenta's protective role against reduced ambient oxygen levels [11]. These milestones highlight the potential of PAI to investigate placental blood distribution and enhance our knowledge of this vital organ [12, 13]. So, it inspired us to assess the physiological state of circadian rhythm-disrupted placental tissue slices with photo-acoustic microscopy. Our findings reveal that photoacoustic imaging/microscopy yields promising results to identify even subtle developmental irregularities in placental growth.

The link between placental development and circadian rhythm was first identified in a groundbreaking study published in 2008, which observed that pregnant rats exposed to constant light produced lighter and smaller placentas [14]. This discovery gained significant attention from the scientific community due to its potential negative influence on subsequent generations. Since then, researchers have explored various aspects of this connection, including its causes, preventive measures, detection methods, and overall significance [15]. For instance, studies have revealed that circadian rhythm disruption during pregnancy can lead to hormonal alterations, impaired fetal growth [16, 17], and an increased risk of non-communicable diseases (NCDs), such as obesity, diabetes, cardiovascular diseases, and specific cancers [18, 19]. Furthermore, investigations have explored the relationship between gestational chrono disruption (GCD), intrauterine growth restriction (IUGR), and the long-term susceptibility of IUGR infants to NCDs.

Despite extensive research into the broader implications and potential outcomes, our understanding of the underlying cellular and molecular pathways remains somewhat incomplete [20, 21]. This limitation is primarily attributed to the complex role of the placenta. It facilitates metabolic gas exchange, eliminates fetal waste, regulates pregnancy-related hormones to modulate the mother's physiological state, and serves as a crucial immune barrier [22]. Several mechanisms are triggered by hormonal activities and continue in parallel to provide as many resources as possible to benefit the growing fetus. In our study, we worked with the mature murine placentas, which are composed of three primary regions: the labyrinth zone (LZ), the spongiotrophoblast (ST) (junctional) zone, and the decidua. The decidua, the maternal side, comprises cells essential in forming the surface vasculature and migrating glycogen cells (GCs) from the junctional zone [23, 24]. The junctional zone houses the placenta's primary endocrine cells—trophoblast giant cells—and temporary GC clusters, vital for fetal development [25, 26]. Notably, only maternal blood vessels are found here [27]. The fetal side, the labyrinth region, contains trophoblast cells, maternal sinuses, and fetal blood arteries, pivotal for maternal-fetal blood exchange and fetal nutrition, emphasizing the importance of tight cellular regulation [28]. In addition to the cell composition, the relative size of these placental zones is of significant importance. The dimensions of these zones directly influence the placenta's capacity for nutrient transfer, thus having profound implications for fetal development. An optimal-sized labyrinth zone, for example, ensures an efficient exchange of oxygen and nutrients between the maternal and fetal circulations. Conversely, deviations in the size of these zones can lead to compromised placental function, potentially impacting fetal growth and overall pregnancy outcomes. Hence, understanding the complex composition and the respective sizes of the mouse placenta zones is crucial in the broader context of mammalian reproductive biology, contributing to insights that may inform interventions for placental dysfunctions and adverse pregnancy outcomes.

Today, we still investigate the possible outcomes of circadian rhythm disruption on growing fetuses due to poor placental development. Current monitoring methods for fetal development usually revolve around regular or Doppler ultrasound scans and some additional serum/blood testing. Even though these methods are proven reliable in the major developmental milestones of fetuses, they undermine minor developmental deficiencies during the course of pregnancy. The present study aims to 1) replace the traditional HE staining with photo-acoustic imaging/microscopy in assessing the physiological state of the placenta concerning changes in the sizes of the primary zones and 2) show a link between the sizes of primary zones and circadian rhythm disruption.

## 2. Materials and Methods

### 2.1. Ethics Statement

Female BALB/c mice, six weeks of age, were obtained from the Animal Research Unit of Akdeniz University, Antalya, Turkey. The experimental procedures were carried out with the explicit approval of the Animal Care and Use Committee of the Faculty of Medicine at Akdeniz University, and they were strictly adherent to the ethical approval protocol (Ethical Approval No. 2023.07.002). Upon the commencement of the experiments, all mice were housed in a controlled environment with a regulated 12-hour light/dark cycle and provided unrestricted access to food. At the end of the experiments; the euthanization of mice was performed through cervical dislocation after receiving anesthesia with a mixture of ketamine and xylazine (0.1 ml/10g body weight, intraperitoneal injection to collect fetus and placenta. Additionally, the weight of the placenta and fetus samples were noted.

### 2.2. Preparation of mouse

Pregnancy in female mice was induced by mating them with fertile males of the same strain. Each cage was allocated one male mouse to mate with two female mice. The appearance of a copulatory plug was considered as day one of pregnancy, resulting in the formation of two distinct groups. In the control group (C), female mice were exposed to a standard 12L-12D light-dark cycle (lights on at 06:00, lights off at 18:00) throughout the gestation period. Placentas and fetuses were collected from this group on the 16th and 18th days of pregnancy. In contrast, the maternal circadian rhythm disrupted group (MCRD) underwent a manipulated light-dark cycle. Starting from the day of vaginal plug detection, the standard 12L-12D cycle was advanced by 6 hours every five days, to mimic the shift work model, deliberately disrupting the circadian rhythm [29]. Similar to the control group, fetuses and placentas were retrieved from the MCRD group on the 16th and 18th days of gestation.

### 2.3. Preparation of tissue sections

Placentas and fetuses were collected from all experimental groups on the 16th and 18th days of gestation. Initially, each fetus and placenta were weighed using precision scales, and their respective weights were documented. The placentas were then processed for subsequent histological analyses to discriminate the placental zones. For these analyses, the placental tissues were fixed in 10% formaldehyde (Merck, Germany) for 24 hours to preserve their structure. Subsequently, they were dehydrated using a graded ethanol series and embedded in paraffin to facilitate sectioning. Sections, five micrometers in thickness, were prepared from each placenta. To maintain standardization, only sections derived from areas passing through the umbilical cord were collected. Prepared tissue sections were scanned under the photo-acoustic microscope.

Following the photo-acoustic microscope (PAM) scans, the paraffin was removed from each tissue section, and they were rehydrated. The sections were then stained with hematoxylin-eosin to highlight cellular and structural details. All stained sections were imaged using a Zeiss (Oberkochen, Germany) light microscope for further evaluation.

*2.4. Photo-acoustic microscope*

The photoacoustic microscopy images were collected with a homemade optically resolved PAM, as seen in Fig. 3. The samples on glass slides were immersed in water. A laser (Spectra-Physics, Explorer One) operating at 532 nm with a repetition rate of 20 kHz and an average power of 10 mW was focused on the samples with a 50 mm lens. Emitted acoustic waves from remaining blood content were received through a 5 MHz single-element ultrasonic transducer and amplified (50dB) with a two-stage amplification scheme in all experiments. To record the photo-acoustic signals coming from the tissue slices, a custom MATLAB (SN: 1076960) code simultaneously translated the motorized stages and recorded the emitted signals through a DAQ card (Gage Compuscope). Later, recorded signals were constructed to the scanned position values to form the photo-acoustic images.

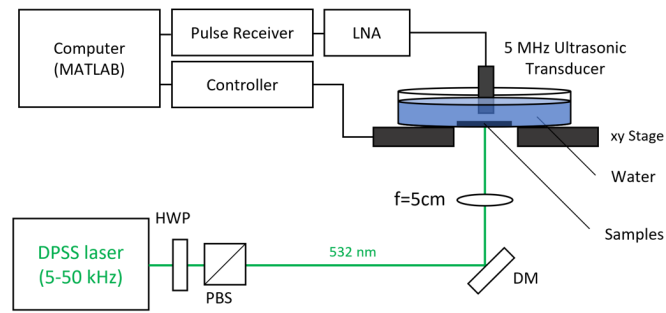

Figure 1. Schematic of the optically resolved photo-acoustic microscope (OR-PAM).

## 3. Results and Discussion

This study explored the feasibility of photo-acoustic microscopy in placental tissue assessment. As a model, we introduced the pathological pregnancy case of circadian rhythm disruption. Our motivation was to demonstrate the sensitivity of PAM in identifying subtle developmental irregularities in placental growth. Methodologically, we compared placental tissue slices collected from circadian rhythm disrupted (MCRD) and undisrupted (control) rats and later imaged them with a photo-acoustic microscope and HE-stained versions with a light microscope.

The murine species characteristically possess a gestational period of around 19-20 days. A critical milestone in embryonic development, initiating the first placental circulation, begins on the 9th day. Subsequent placental development spans from this day until approximately the 15th, reaching maturation by the 16th day. Consequently, we elected to examine the placental samples at days 16 and 18 of development to juxtapose the vascular structure of the placenta after the completion of maturation and proximate to the birth [21].

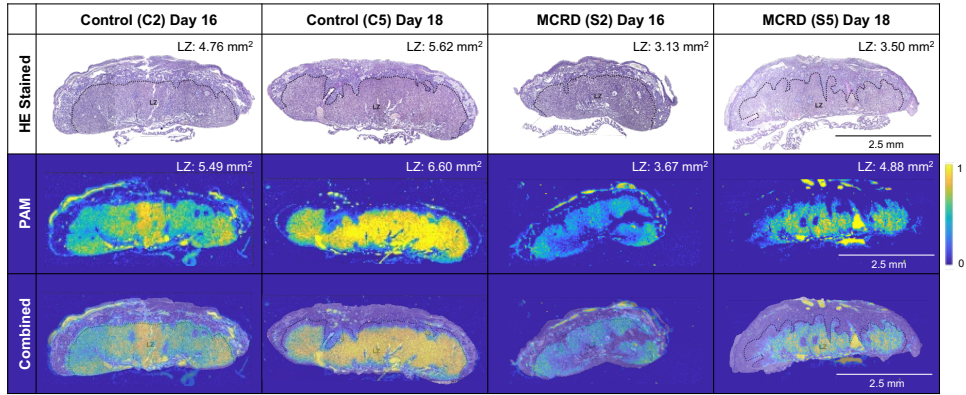

Figure 2. HE stained and PAM images for C2 and S2 for Day 16; C5 and S5 for Day 18 from gestation.

First, we performed PAM experiments with the setup shown in Figure 1, and later, we applied the staining procedure and recorded the HE-stained images. The acquired PAM images (2nd row), along with the corresponding HE-stained images (1st row), are displayed in Figure 2. In reproductive medicine, the labyrinth zone is known for its high blood content due to maternal exchange, and it dramatically influences the health of the growing fetus [25, 30]. So, in general practice, traditional HE-staining procedures are adopted to reveal the LZ to explore the role of placental zones in growing fetuses. Because the only absorber at the laser wavelength of 532 nm in the tissue composition is red blood cells (RBCs), the contrast observed in the PAM images arises from the remaining residual RBCs within LZ [2]. As a result, when we look carefully at the combined overlays on the 3rd row, we can easily observe a nice overlap in LZ borders between PAM and HE-stained images. These overlapped images suggest PAM is a suitable imaging alternative to the traditional HE staining process.

Table 1: Calculated area values acquired from PAM images for the labyrinth zone, placenta, and ratio of these values.

| Days | # | Control | | | # | MCRD | | |
| --- | --- | --- | --- | --- | --- | --- | --- | --- |
| | | PAM Labyrinth (mm$^2$) | HE Labyrinth (mm$^2$) | Placenta (mm$^2$) | | PAM Labyrinth (mm$^2$) | HE Labyrinth (mm$^2$) | Placenta (mm$^2$) |
| 16 | C1 | 3.74 | 4.45 | 6.94 | S1 | 1.72 | 2.52 | 4.50 |
| 16 | C2 | 4.76 | 5.52 | 8.33 | S2 | 3.13 | 3.94 | 5.53 |
| 16 | C3 | 4.30 | 5.06 | 8.41 | S3 | 3.02 | 4.28 | 6.02 |
| | Mean: | 4.26 | 5.01 | 7.89 | Mean: | 2.62 | 3.58 | 5.35 |
| 18 | C4 | 4.80 | 6.53 | 7.75 | S4 | 2.88 | 5.70 | 6.65 |
| 18 | C5 | 5.62 | 7.74 | 8.54 | S5 | 3.50 | 4.80 | 6.94 |
| 18 | C6 | 5.08 | 5.17 | 7.76 | S6 | 2.69 | 3.32 | 5.88 |
| | Mean: | 5.16 | 6.48 | 8.02 | Mean: | 3.03 | 4.61 | 6.49 |

Later, to further emphasize the sensitivity of PAM to physiological clues, we calculated the respective areas of the entire placental tissue and the LZ, as indicated by the PAM signals and HE-stained counterparts. The values are tabulated in Table 1 for both modalities. As the gestation progresses from day 16 to 18 for control animals, the mean placenta area slightly increases (<1.65%), but the mean LZ area increases by 29%-HE (21%-PAM) to keep up with the nutrition demand of the growing fetus. In other words, the mean LZ area to total placenta ratio increased from 63%-HE (54%-PAM) for day 16 to 80%-HE (64%-PAM) for day 18. For circadian rhythm-disrupted animals, the placental area from day 16 to 18 increased by 21% as opposed to the control case, trying to compensate for the negative impact of underdevelopment.

The mean LZ area to total placenta ratio did not increase at the same pace as control samples. Day 16: 66%-HE (49%-PAM) and day 18: 66%-HE (47%-PAM). When we compare the size on days 16 and 18 to control, all absolute area values measured for MCRD are at least 0.68 and 0.8 times smaller, respectively. Such departure suggests potential developmental implications of disrupted circadian rhythm, particularly affecting the LZ and placental growth. These findings emphasize the significance of proper circadian regulation during gestation and its potential impact on the development of the placenta, with specific relevance to the LZ region.

To complete the link between fetus health and placental development, we measured the placenta and fetus weights to provide additional evidence supporting the correlation between the PAM images and the growth status of fetuses. Figure 3 depicts the placental and fetal weights of control and MCRD group animals, revealing the occurrence of IUGR as a direct consequence of MCRD. The impact of MCRD is evident in the reduced placental and fetal weights observed on days 16 and 18 of pregnancy. The lack of LZ development is implicated as the underlying cause of underweight births in mice. Furthermore, the fetal/placental weights and PAM data for MCRD mice on the 18th day indicate that as the pregnancy progresses, the placenta attempts to compensate for the underweight fetus. But still, compared to control animals on day 18, MCRD fetuses are 0.71 times smaller. The data presented in Figure 3 indicates that despite the placental compensatory effort, the weights measured for the control fetuses were not attained in the MCRD group. In other words, the placental compensation was insufficient to fully restore the weights of the MCRD-affected fetuses to the level observed in the control group.

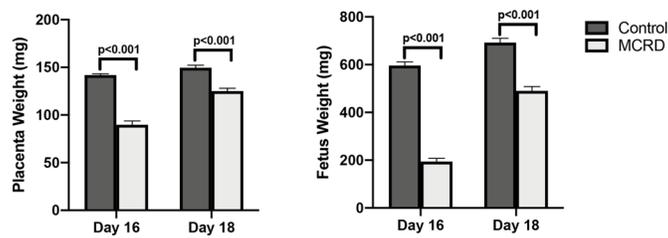

Figure 3: Weight data of placenta and fetus gathered from control and MCRD mice (N=6).

PAI's unique capability to visualize blood-rich areas, such as the LZ, renders it a crucial tool in comprehending placental pathologies. Disruptions in placental vasculature can result in complications like IUGR or preeclampsia, and PAI's precise imaging offers the promise of early detection and effective management of these conditions. Furthermore, the safety and high resolution of PAI highlight its future potential to advance the study of placental disorders and improve perinatal outcomes.

Our study has provided valuable insights into placental development, but certain limitations should be acknowledged. Our methodology is based on paraffin sectioning and 2D imaging. While this 2D approach has proven effective in our study—thanks to using and evaluating HE-stained sections, which have verified the successful capture of the placental zones via PAI—it inherently cannot provide a comprehensive 3D representation of placental architecture. Moreover, tissue fixation, which is a prerequisite for our methodology, introduces an extra time-consuming step and may potentially alter the precision of our results. Finally, the lack of calibration for blood imaging may have resulted in a less precise quantification of blood-rich areas such as the LZ. Despite the aforementioned limitations, it's important to emphasize that the utilization of PAI, especially for evaluating unlabeled tissue morphology in healthy and pathological conditions, holds significant promise. This is particularly relevant for studies focusing on the vasculature, given PAI's unique ability to visualize and quantify blood-rich areas non-invasively. Even with the current constraints, the successful application of PAI in our study underlines its potential as an effective tool in biomedical research. As we refine the technology and methods, we anticipate that PAI will be increasingly instrumental in advancing our understanding of complex biological structures and processes, opening new frontiers in the diagnosis and treatment of various vascular-related health conditions. The broader implication of this research is that it highlights the promising role of photoacoustic imaging in advancing our understanding of placental health in such affected individuals. We thoroughly examined the notable variations in the development of placental tissue between control mice and those exposed to MCRD.

## 4. Conclusion

Currently, the conventional HE staining procedure is widely preferred to separate the essential zones in placental tissue evaluation for mice. By introducing the PAM modality, 1) we increased the sensitivity in separating blood-rich labyrinth zones and 2) dismissed the timeconsuming HE staining process. Such effortless zone marking capability of PAM and a strong link between the health of the fetus/placenta and the relative size of the labyrinth zone will introduce a new battlefield to PAM applications in placenta-related research. With this work, we are also introducing a new pathological case of circadian disruption to PAM, which will significantly enhance the difficulties for histology scientists willing to evaluate the essential zones of the placenta.

The broader implications of our findings extend beyond placental health, contributing to the broader field of medical research and potentially influencing clinical practices to improve maternal and fetal outcomes. The improved ability to evaluate pathological models without labeling is a significant advancement. This reduces time, resources, and potential artifacts. Furthermore, PAI's capacity to assess fresh tissue, especially small-volume samples, widens its potential applications. This ability is especially valuable in clinical settings where timely analysis is essential, enabling real-time assessment of biopsy samples for rapid diagnosis and intervention. The ability to differentiate placental zones based on their red blood cell content provides a detailed comprehension of placental structure and function, which can be pivotal in diagnosing and managing conditions impacting placental blood flow and fetal nutrition, such as IUGR or preeclampsia.

## 5. Back matter

### 5.1 Funding

Multiple organizations supported this work. CCO (the corresponding author) received funding from The Scientific and Technological Council of Turkey (TUBITAK) [grant number: 119S121] for this work. The photo-acoustic microscope was supported by TUBITAK under grant 119F319, and the State Planning Organization of Turkey (DPT) under grant 2009K120520. Alireza Khoshzaban acknowledges funding from the European Union Horizon 2020 research and innovation programme under the Marie Skłodowska-Curie grant agreement No 812780 (ActiveMatter). The funders had no role in study design, data collection and analysis, the decision to publish, or the preparation of the manuscript. The authors gratefully acknowledge the use of the services and facilities of the Koç University Research Center for Translational Medicine (KUTTAM), funded by the Presidency of Turkey, Presidency of Strategy and Budget.### 5.2 Disclosures

The authors declare no conflicts of interest.

## 6. References

[1] A.B.E. Attia, G. Balasundaram, M. Moothanchery, U.S. Dinish, R.Z. Bi, V. Ntziachristos, M. Olivo, A review of